\documentclass[prd,twocolumn,floatfix,amsmath,nofootinbib,amssymb,floatfix]{revtex4}
\usepackage{graphicx,color,dcolumn,booktabs,bm}
\usepackage{longtable,lscape}
\usepackage{pdfpages}
\usepackage{txfonts}
\usepackage{overpic}
\usepackage{amssymb}
\usepackage{makecell}
\usepackage{indentfirst}
\usepackage{feynmf}   
\usepackage{slashed}  
\usepackage{cases}
\usepackage{color}
\usepackage{multirow}
\usepackage{threeparttable}
\usepackage{epstopdf}
\usepackage{enumerate}
\usepackage{subfigure}
\usepackage{diagbox}
\usepackage{graphicx,color,dcolumn,booktabs,bm}
\usepackage{mathrsfs}
\usepackage{cancel}
\usepackage{float}
\usepackage[colorlinks,
            citecolor=blue,
            anchorcolor=red,
            menucolor=red,
            linkcolor=red,
            filecolor=red,
            runcolor=red,
            urlcolor=blue,
            frenchlinks=red]{hyperref}

\begin{document}
\title{Updated $\Sigma_{c}$ excitation spectrum including the $2P$- and $1F$-wave fine structures}
\author{Zhen-Yu Li$^{1}$}
\email{zhenyvli@163.com }
\author{Guo-Liang Yu$^{2}$}
\email{yuguoliang2011@163.com }
\author{Zhi-Gang Wang$^{2}$ }
\email{zgwang@aliyun.com }
\author{Jian-Zhong Gu$^{3}$ }
\email{gujianzhong2000@aliyun.com }

\affiliation{$^1$ School of Physics and Electronic Science, Guizhou Education University, Guiyang 550018,
China\\$^2$ Department of Mathematics and Physics, North China Electric Power University, Baoding 071003,
China\\$^3$ China Institute of Atomic Energy, Beijing 102413,China}
\date{\today }

\begin{abstract}
In order to study the new $\Sigma_{c}$ baryons observed by the LHCb collaboration recently, we update the previously predicted $\Sigma_{c}$ excitation spectrum by extending the excited states up to the highly excited $2P$- and $1F$-wave states. The calculation is performed in the framework of the relativized quark model and Gaussian expansion method by considering the mixing effect. The radially and orbitally excited states in the calculation are determined by the truncation convention proposed in this work and the mechanism of the heavy-quark dominance. The contribution of each Hamiltonian term to the energy levels is precisely calculated by using the two-step Gaussian expansion method. Then, the concerned $\Sigma_{c}$ states are obtained and the $\Sigma_{c}$ excitation spectrum including the $2P$- and $1F$-wave fine structures is established. Based on these, several important topics are discussed and analyzed, including the root-mean-square radii, off-diagonal matrix elements, mixing effect, fine structures and assignments of the newly observed $\Sigma_{c}(2800)^{0}$, $\Sigma_{c}(2900)^{0}$ and $\Sigma_{c}(3200)^{0}$ states. Meanwhile, it is found that the approximate decoupling phenomenon emerges in different $nL$-wave groups, here, $n$ is the radial quantum number and $L$ is the orbital angular momentum quantum number. This implies the reasonability of the truncation convention and is of great significance for further improving the calculation scheme for all the baryon spectra. Moreover, it is demonstrated that in the very high-lying $nL$-wave groups such as the $1F$-wave group, the energy level gaps become small even degenerate, and each of the baryon states has a big radius and therefore a loosely bound structure. In addition, by examining the experimental and calculated data, we suggest more precise experiments to search for more $\Sigma_{c}$ baryon states in the predicted $1P$- and $2P$-wave fine structures whose central energies are around 2822 and 3210 MeV, respectively.

\end{abstract}
\maketitle
\section{Introduction}\label{sec1}
The heavy baryon spectroscopy has made a significant progress and greatly facilitated the study of the strong interaction dominated by non-perturbative QCD~\cite{F101,CHY15,P251,F411,Chenhx23,Crede24}. On the experimental side, many singly heavy baryon states have been measured in recent decades~\cite{F201,F2021,P2940,F205,Babar06,P2910,P2882,F206,F207,F208,LHCb2002,F210,F209,F211,F212,F213,F4401,F4402,F4403,F203,F204,F214,LHCb25,LHCb26,LHCb262}. In the new Review of Particle Physics (RPP) by the Particle Data Group (PDG), more than 70 singly heavy baryons were collected~\cite{F201}. The rudimentary spectral structure of low-lying singly heavy baryons was constructed. In particular, the observed fine structures of low-lying negative-parity heavy baryon states have advanced the study of the strong interaction within the baryon systems, such as the $\{\Omega_{c}(3000)^{0}$, $\Omega_{c}(3050)^{0}$, $\Omega_{c}(3065)^{0}$, $\Omega_{c}(3090)^{0}$, $\Omega_{c}(3120)^{0}$\} and $\{\Omega_{b}(6316)^{-}$, $\Omega_{b}(6330)^{-}$, $\Omega_{b}(6340)^{-}$, $\Omega_{b}(6350)^{-}\}$.

On the theoretical side, the singly heavy baryon spectra have been systematically studied by various theoretical approaches, including lattice QCD~\cite{MP17,HB20,Chen23}, QCD sum rules~\cite{P21,P22,P221,F315,f17,f17p8,f17p7}, effective field theory~\cite{F304,F303}, chiral perturbation theory~\cite{P23,P24}, Regge trajectories~\cite{P25,CJK24} and quark potential models~\cite{CB15,F403,F405,RNF22,YG18,WKL19,WKL22,f17p13,f17p10,F601,zhao24,F315p,F4071,f17p14,f17p11}. Especially, some important properties in the heavy baryon spectra were revealed, such as the heavy quark symmetry~\cite{Isgur91}, heavy quark limit~\cite{GG90} and heavy quark effective theory~\cite{F403,EE90}. And some approximation methods were adopted in the study, among which the diquark approximation was particularly noteworthy~\cite{MA93}. All theoretical efforts contribute to a deeper understanding of the nature of the heavy baryon spectral structure.

As an indispensable quark potential model~\cite{F306}, the relativized quark model (RQM) was first used to describe baryons in 1986~\cite{F401,F402}. Over 40 years, the RQM has undergone a significant development in the study of the heavy baryon spectroscopy in the following two main aspects: (a) the calculation scheme and (b) the high-precision calculation, which was a severe challenge for all the theories and models including the RQM. In recent years, the orbital excitation mechanism in heavy baryons has been gradually clarified~\cite{F403,YO15,Isgur81,Chen19} and the mechanism of the heavy-quark dominance (HQD) was proposed which determines the spectral structure of low-lying singly and doubly heavy baryons~\cite{Li241,Li242}. This provides a strong support for the improved calculation scheme as discussed below.

The Gaussian expansion method (GEM)~\cite{Ka88,Hi03} has been introduced in the study of the hadron spectroscopy for decades, which can perform high-precision calculations for multi-quark systems~\cite{YG18,YO15,F601,Luo23,Luo25}. In order to further enhance the calculation power of the GEM, the two-step GEM was proposed early this year~\cite{Li26}, which can work out all kinds of the complicated matrix elements and overcomes the calculation difficulty that had plagued the RQM over 40 years~\cite{Li26,zhao24}. Then, an improved version of the calculation scheme has been established which is actually a dedicated combination method for low-lying heavy baryons in the frame work of the RQM and GEM by considering the HQD and mixing effect~\cite{Li26,zhao24,Li251,Li252}. An important feature of the improved calculation scheme is that the final state obtained is a baryon state labeled with (mass)$J^{P}$ after taking into account the mixing effect, where the orbital angular momentum quantum number $L$ is no longer a good quantum number.
Based on these progresses, the RQM has been successfully applied to explain the fine structure and excitation spectra of low-lying singly heavy baryons with a very high-precision~\cite{Li26,Li262} .

Very recently, the LHCb collaboration observed the known $\Sigma_{c}(2455)^{0}$, $\Sigma_{c}(2520)^{0}$ and $\Sigma_{c}(2800)^{0}$ baryons, along with two new states: $\Sigma_{c}(2900)^{0}$ and $\Sigma_{c}(3200)^{0}$, in the $\Lambda_{c}^{+}\pi^{-}$ invariant-mass spectrum by analyzing the $B^{-}$ $\rightarrow$ $\Lambda_{c}^{+}\bar{p}\pi^{-}$ decay~\cite{LHCb262}. This result expands the landscape of charm baryon spectroscopy. The two new states $\Sigma_{c}(2900)^{0}$ and $\Sigma_{c}(3200)^{0}$ have attracted  theoretical attention~\cite{Luo26,Zhang26,Hiyama26,Chen26}. Our preliminary calculations indicate that the $\Sigma_{c}(3200)^{0}$ enters the energy level region of the $2P$-wave states, which exceeds the energy range of the low-lying $\Sigma_{c}$ baryons as discussed in Ref.~\cite{Li262}. Meanwhile, there should be a fine structure in the energy region of these $2P$-wave states, namely the $2P$-wave fine structure. Therefore, these new baryons are crucial for the understanding of the charm baryon structure and the specific feature of the $\Sigma_{c}$ spectrum. And, it is a very interesting topic to explore this new fine structure, update the $\Sigma_{c}$ spectrum and give a reasonable explanation for these newly observed $\Sigma_{c}$ states. This is precisely the main purpose of this work.

The remainder of this paper is organized as follows. In
Sec.~\ref{sec2}, the calculation strategy used in this work is introduced, including the truncation convention of higher excited states in the RQM, calculation scheme for the low-lying states and calculation scheme used in this work. Sec.~\ref{sec3} presents the calculated results and the relevant discussions including the root-mean-square (rms) radii, off-diagonal matrix elements, mixing effect, updated $\Sigma_{c}$ spectrum, fine structures and assignments of the newly observed $\Sigma_{c}$ baryons. And Sec.~\ref{sec4} is reserved for the concluding remarks.

\section{The calculation strategy}\label{sec2}

The calculation strategy for the excitation spectrum is based on the properties of the spectral structure and its compatibility with the experimental observation.  Based on the experimental observation~\cite{F201} and our previous theoretical study~\cite{Li241,Li262}, there are four specific and reliable properties of the singly heavy baryons: (1) The observed heavy baryons predominantly populate the low-lying excited states and their total number remains small; (2) The orbital excitation can be explained by the HQD mechanism and the predicted low-lying baryon spectra can reproduce the current experimental data nicely; (3) Most of the low-lying baryon states can be explained within the $1S$-, $2S$-, $3S$-, $1P$- and $1D$-wave states by considering the mixing effect; (4) The calculation indicates that the $\Sigma_{c}(3200)^{0}$ is the only high-lying excited state which exceeds the range of the above low-lying baryon states. According to these properties, the corresponding calculation strategy is given as follows.

1) Truncation convention of highly excited states in the RQM

Due to the linear confinement potential, the RQM can generate infinite highly excited energy levels in theory. However, the observed states are always rather limited in practice. Therefore, the truncation of the excited states is necessary to be made in the calculation. But, there is no universal consensus on the criteria for the truncation. Here, we adopt the following truncation convention of highly excited states in the RQM.

All of the radially (labeled with $n$) and orbitally (labeled with $L$) excited states with the same $n$ and $L$ values form a group named as the ``$nL$-wave group'' in this work, such as the $1P$-wave group. In each $nL$-wave group, the excited states come from various excitation modes such as the $1P_{\lambda}$-wave states coming from the $\lambda$-mode as discussed below.

The truncation is performed not only for those high-lying $nL$-wave groups but also for the states within a group. First, those $nL$-wave groups with higher excitation energies should be excluded as much as possible, unless they are necessary for explaining the experimental data. Then, within each included (or remained) $nL$-wave group, only the excitation mode with a lower excitation energy is generally retained for calculations, which is coined as the HQD mechanism~\cite{Li241}.

The truncation convention adopted in this work effectively reduces the model space dimension used in calculations. It is different from that used in the 1986 paper~\cite{F402} and needs to be tested by theory and experiment.

2) The calculation scheme for the low-lying states

For the low-lying singly heavy baryon states, a dedicated calculation scheme, namely the improved calculation scheme, has been well established~\cite{Li26,Li262} after a long period of exploration as discussed in Sec.~\ref{sec1} and is different from that used in the 1986 paper~\cite{F402}. Briefly, the improved calculation scheme performs the calculation of the low-lying baryon states in the framework of the RQM and GEM, in which the HQD is used to determine which orbitally excited states should be included in the calculation and the mixing effect is considered to obtain the baryon states (the physical observed states). In all the low-lying singly heavy baryon families, the included radially and orbitally excited states ($nL$) are the $1S$-, $2S$-, $3S$-, $1P_{\lambda}$-, $1P_{\rho}$- (only for the charm baryons) and $1D_{\lambda}$-wave states, which are sufficient to cover all the observed low-lying singly heavy baryon states as discussed in our previous work~\cite{Li262}. More details of the improved calculation scheme can be found in Ref.~\cite{Li262}.

3) The calculation scheme used in this work

For describing the $\Sigma_{c}(3200)^{0}$, we should take into account the $2P$-wave states. Safely, we should also consider the influence of the $1F$-wave states, because their energy levels are close to those of the $2P$-wave states. Thus, the concerned negative-parity baryon states should be obtained from the $1P$-, $2P$- and $1F$-wave states by considering their mixing effect. The positive-parity baryon states do not need to be updated at present since they can be directly moved over from Ref.~\cite{Li262}. The calculation scheme in this work is as follows.

\begin{figure}[htbp]
\centering
\includegraphics[width=8.5cm]{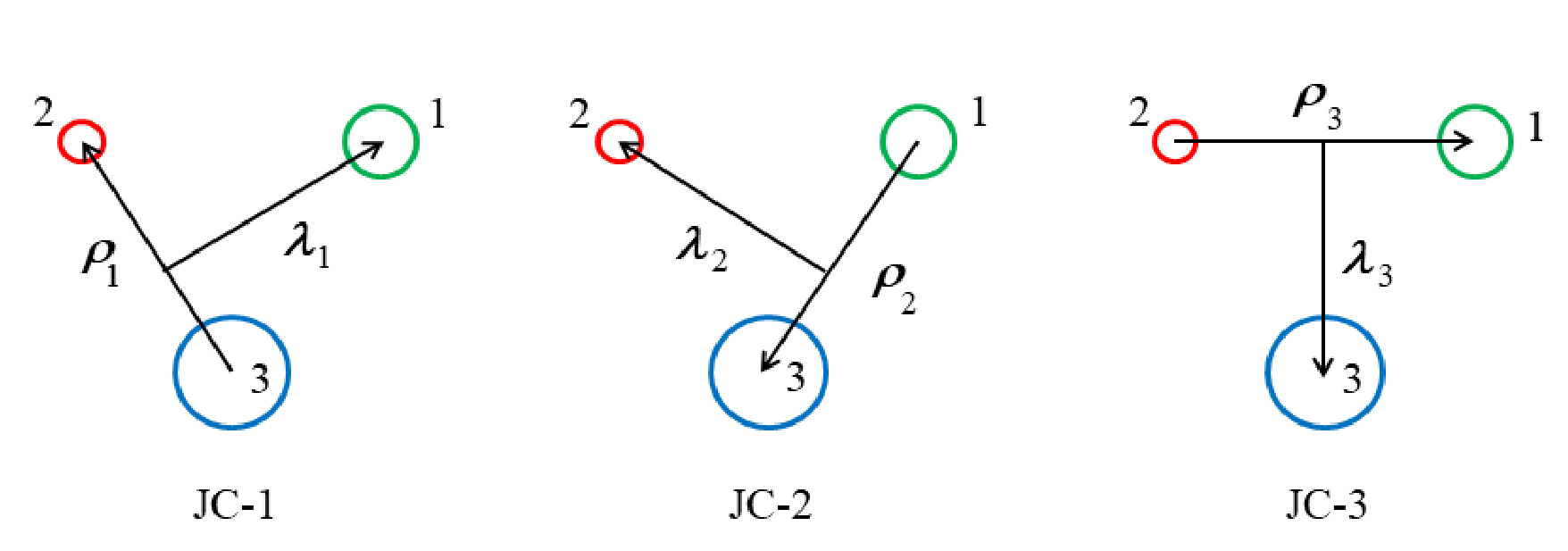}
\caption{There are three channels of the Jacobi coordinates for a three-quark system, labeled with $(\boldsymbol\rho_{k}$, $\boldsymbol\lambda_{k})$ ($k$=1, 2, 3). The channel 3 (JC-3) is selected for defining the spin and orbital wave functions in the singly heavy baryon system. All the quarks are numbered for ease of calculation, and the third quark refers specifically to the heavy quark.}
\label{fig1}
\end{figure}

First, the total wave function of a singly heavy baryon is expressed as the direct product of the color, flavor and spin-orbit wave functions. Assuming the two light quarks conform to the flavor $SU(3)$ symmetry, the $\Sigma_{c}$ baryons belong to the sextet ($\mathbf{6}_{F}$) of the flavor symmetric states. Then, the spin-orbital wave functions are defined in the special Jacobi coordinates JC-3 with $(\boldsymbol\rho_{3}$, $\boldsymbol\lambda_{3})$ $\equiv$ $(\boldsymbol\rho$, $\boldsymbol\lambda)$ (see Fig.~\ref{fig1}) and have the following coupling scheme~\cite{F403},
\begin{eqnarray}
|(J^{P})_{j},L\rangle = |\{[(l_{\rho} l_{\lambda} )_{L}(s_{1}s_{2})_{s_{12}}]_{j} s_{3}\}_{J }\rangle,
\label{eq1}
\end{eqnarray}
with parity $P=(-1)^{l_{\rho}+l_{\lambda}}$ and $(-1)^{l_{\rho}+s_{12}}=-1$ due to the total antisymmetry for the two light quarks. $s_{i}$, $l_{\rho(\lambda)}$ and $L$ are the quantum numbers of quark spin, orbital angular momentum (only for the JC-3 of Fig.~\ref{fig1}) and total orbital angular momentum, respectively. Simply, an orbitally excited state is labeled with $(l_{\rho},l_{\lambda})L(J^{P})_{j}$.
$(l_{\rho},l_{\lambda})L$ denotes the orbital excitation mode. It is commonly known as the $\lambda$-mode ($(0,l_{\lambda})L(J^{P})_{j}$ $\equiv$ $L_{\lambda}(J^{P})_{j}$) or the $\rho$-mode ($(l_{\rho},0)L(J^{P})_{j}$ $\equiv$ $L_{\rho}(J^{P})_{j}$) in the literature~\cite{F101}.

Based on the above calculation strategy and HQD~\cite{Li241}, the included orbitally excited states $|\alpha\rangle$ for the negative-parity $\Sigma_{c}$ baryons consist of seven $P$-wave states  $\{|P_{\lambda}(\frac{1}{2}^{-})_{0}\rangle$, $|P_{\lambda}(\frac{1}{2}^{-})_{1}\rangle$, $|P_{\lambda}(\frac{3}{2}^{-})_{1}\rangle$, $|P_{\lambda}(\frac{3}{2}^{-})_{2}\rangle$, $|P_{\lambda}(\frac{5}{2}^{-})_{2}\rangle$, $|P_{\rho}(\frac{1}{2}^{-})_{1}\rangle$, $|P_{\rho}(\frac{3}{2}^{-})_{1}\rangle\}$ and six $F$-wave states $\{|F_{\lambda}(\frac{3}{2}^{-})_{2}\rangle$, $|F_{\lambda}(\frac{5}{2}^{-})_{2}\rangle$, $|F_{\lambda}(\frac{5}{2}^{-})_{3}\rangle$, $|F_{\lambda}(\frac{7}{2}^{-})_{3}\rangle$, $|F_{\lambda}(\frac{7}{2}^{-})_{4}\rangle$, $|F_{\lambda}(\frac{9}{2}^{-})_{4}\rangle\}$.

Second, the Hamiltonian expectation values $\langle H\rangle$ $\equiv$ $\langle \alpha|\hat{H}|\alpha\rangle$/$\langle \alpha|\alpha\rangle$ are calculated in the framework of the GEM.  In the RQM, the Hamiltonian reads~\cite{F402}
\begin{eqnarray}
\notag
\hat{H} &&=\hat{H}^{spin-independent}+\hat{H}^{spin-dependent}\\
         &&\equiv\hat{H}^{mode}+\hat{H}^{s-d},
\label{eq2}
\end{eqnarray}
with
\begin{eqnarray}
\notag
\hat{H}^{mode}=\sum_{i=1}^{3}\sqrt{p_{i}^{2}+m_{i}^{2}}+\sum _{i<j}\hat{H}^{conf}_{ij},\\
\hat{H}^{s-d}=\sum _{i<j}(\hat{H}^{tens}_{ij}+\hat{H}^{cont}_{ij}+\hat{H}^{so(v)}_{ij}+\hat{H}^{so(s)}_{ij}).
\label{eq3}
\end{eqnarray}
It has been confirmed that $\hat{H}^{mode}$ determines the energy levels of the orbital excitation modes $(l_{\rho},l_{\lambda})L$. While, these spin-dependent terms $\hat{H}^{s-d}$ contribute to the energy level shifts and splittings~\cite{Li251} as well as the mixing effect~\cite{Li26}. In calculations, the spin-orbit terms $\hat{H}^{so}$ are commonly divided into the two-body spin-orbit terms $\hat{H}^{so2B}$ and the three-body spin-orbit terms $\hat{H}^{so3B}$ so as to distinguish their contributions to the energy levels and mixing effect.

The above orbitally excited state $|\alpha\rangle$ can be expanded in the generalized Gaussian basis functions $|(\xi,\alpha)^{G}\rangle$~\cite{Li251} with $\xi$ = 1, $\cdot\cdot\cdot$, $n_{max}$.
And the Hamiltonian $\hat{H}$ is represented as a matrix $H$ with $H_{\xi\xi'}$ = $\langle(\xi,\alpha)^{G}|\hat{H}|(\xi',\alpha)^{G}\rangle$. Its solution belongs to a generalized matrix eigenvalue problem
\begin{eqnarray}
\begin{aligned}
\sum^{n_{max}}_{\xi'=1}(H_{\xi\xi'}-EN_{\xi\xi'})C_{\xi'}=0.
\label{eq4}
\end{aligned}
\end{eqnarray}
Here, $N_{\xi\xi'}$ comes from the non-orthogonality of the basis functions. Then, the eigenvalue (or the energy level of the radially and orbitally excited state $|n,\alpha\rangle$) $E_{n\alpha}$ and the eigenvector $\{C_{\xi'}\}_{n\alpha}$ are obtained for the $|n,\alpha\rangle$ state. Here, $n$ represents the radial quantum number. In this work, we calculated the energy levels and their eigenvectors of the included $1P_{\lambda}$-, $1P_{\rho}$-, $2P_{\lambda}$-, $2P_{\rho}$-, and $1F_{\lambda}$-wave states.

At last, the baryon states are obtained by taking into account the mixing effect (or the configuration mixing) between these included excited states. The included states with the same spin-parity values form a $J^{P}$ subspace $\{|m,\alpha\rangle$, $|n,\beta\rangle$, $\cdot\cdot\cdot\}$. Thus, the Hamiltonian $\hat{H}$ becomes a matrix $\tilde{H}$ in this subspace with the diagonal elements $\tilde{H}_{m\alpha,m\alpha}$ = $\langle m,\alpha|\hat{H}|m,\alpha\rangle$/$\langle m,\alpha|m,\alpha\rangle$ and the off-diagonal elements $\tilde{H}_{m\alpha,n\beta}$ = $\langle m,\alpha|\hat{H}|n,\beta\rangle$/$\sqrt{\langle m,\alpha|m,\alpha\rangle\langle n,\beta|n,\beta\rangle}$ ($|m,\alpha\rangle$ $\neq$ $|n,\beta\rangle$)~\cite{Li26}. By diagonalizing the matrix $\tilde{H}$, the baryon states together with their energy levels $(mass)J^{P}$ in this subspace are obtained. For overcoming the difficulty in some complicated calculations of matrix elements and achieving the calculation of all the off-diagonal matrix elements with a high-precision, the two-step GEM is applied in the calculation~\cite{Li26}. Note that the calculation of the matrix elements of the tensor terms $\hat{H}^{tens}_{ij}$ and the three-body spin-orbit terms $\hat{H}^{so3B}_{ij}$ is impossible without the two-step GEM. More calculation details can be found in our previous work~\cite{Li251,Li26,Li262}.

\section{Results and discussion}\label{sec3}

\begin{figure*}[htbp]
\centering
\includegraphics[width=16.5cm, trim=1.5cm 7cm 6cm 1.5cm, angle=0]{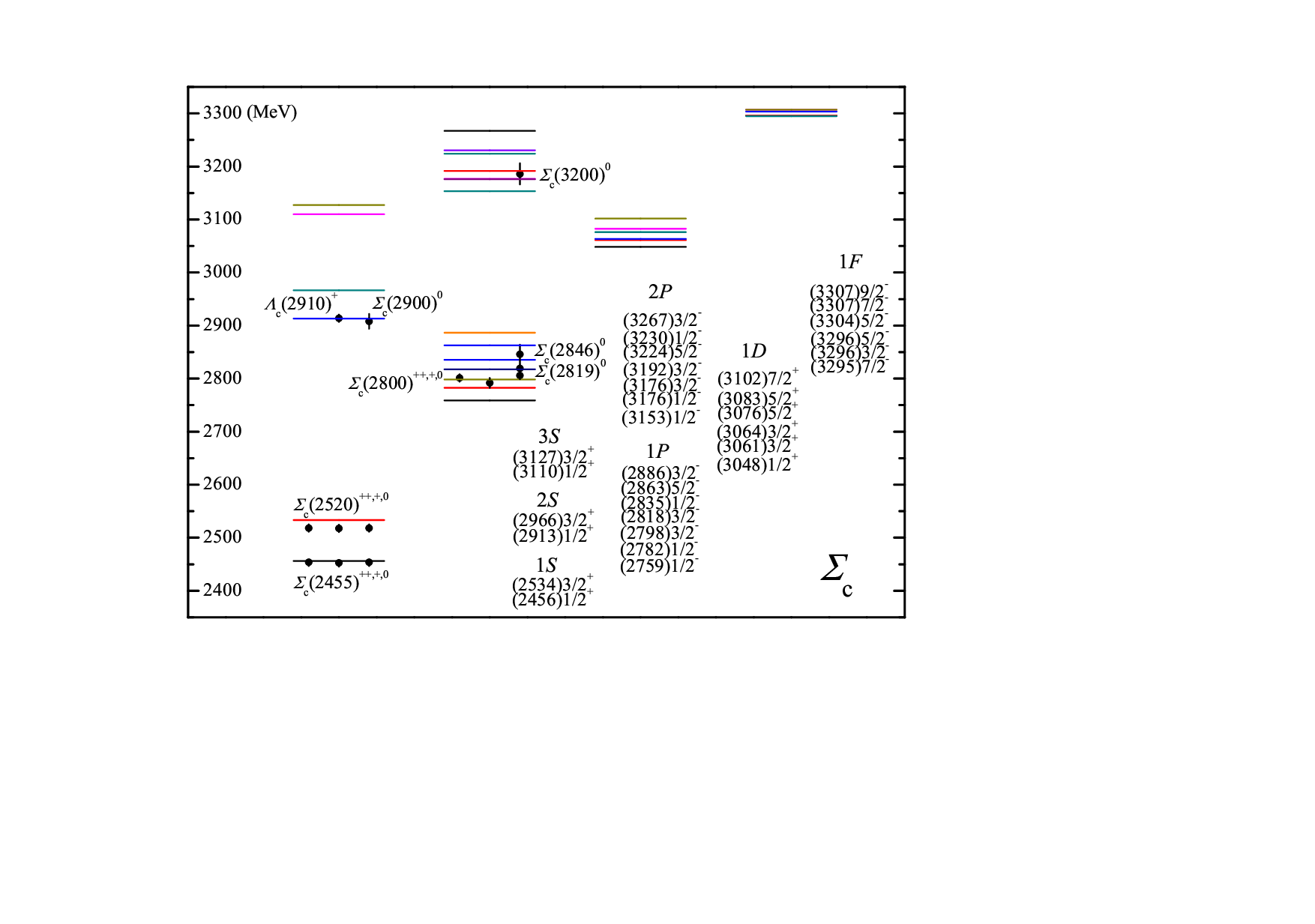}
\caption{Updated spectrum of the $\Sigma_{c}$ baryons including the contribution from the 2$P$- and 1$F$-wave states. The relevant experimental data are also plotted~\cite{F201,F209,LHCb262}. The predicted energy levels are indicated by the horizontal lines. The baryon states are labeled with (mass)$J^{P}$ and arranged in the lower right corner, being divided into seven groups which approximately correspond to the $1S$-, $2S$-, $3S$-, $1P$-, $2P$-, $1D$- and $1F$-wave groups, respectively. }
\label{fig2}
\end{figure*}

Based on the above calculation scheme, we analyzed the included radially and orbitally excited states $nL(J^{P})_{j}$. Their energy levels and root-mean-square (rms) radii ($\langle r^{2}_{\rho}\rangle^{1/2}$ and $\langle r^{2}_{\lambda}\rangle^{1/2}$ ) are presented in Table~\ref{tb1}, where the geometric structure and excitation feature of the $\Sigma_{c}$ family are demonstrated. For examining the mixing effect between the negative-parity excited states, we calculated the off-diagonal matrix elements in the $\frac{1}{2}^{-}$, $\frac{3}{2}^{-}$, $\frac{5}{2}^{-}$ and $\frac{7}{2}^{-}$ subspaces, respectively, as shown in Table~\ref{tb2}. Then, we present the Hamiltonian matrix in each $J^{P}$ subspace and give its eigenvalues and eigenvectors when including (see Table~\ref{tb3}) and excluding (see Table~\ref{tb4}) the $1F_{\lambda}$-wave states, respectively. For the sake of completeness and comparison, Fig.~\ref{fig2} demonstrates the results of the negative-parity baryon states in Table~\ref{tb3} and those of the positive-parity baryon states from Ref.~\cite{Li262}. They constitute the updated mass spectrum of the $\Sigma_{c}$ family as shown in Fig.~\ref{fig2}, where the experimental data are also plotted for the convenience of analysis. With the calculated results, we discuss the following important issues.

1) Root-mean-square radii and excitation energy levels

Both of the root-mean-square (rms) radii and excitation energy levels are influenced by all of the quantum numbers $n$, $L$, $J$, $P$ and $j$ which characterize the included radially and orbitally excited states as shown in Table~\ref{tb1}. From Table~\ref{tb1}, one can see that $\langle r^{2}_{\lambda}\rangle^{1/2}$ and $\langle r^{2}_{\rho}\rangle^{1/2}$ are strongly influenced by $n$ and $L$ which determine the energy level structure and less affected by $J$ and $j$, so are the excited energy levels. The parity $P$ is completely determined by the orbital momentum $L$ and therefore strongly affects the radii and excited energy levels. As $n$ and $L$ increase, the rms radii and excited energy levels go up obviously. The $S$-wave states serve as an example to demonstrate the strong dependence of the rms radii and excited energy levels on the radial quantum number $n$. With the increase of $n$, the rms radii and excited energy levels increase significantly for the $S$-wave states. In the case of the same $n$, $L$, $J$, $P$ and $j$, the $\rho$-mode and $\lambda$-mode make the difference in the rms radii and excited energy levels as shown in the $1P$- and $2P$-wave states in Table~\ref{tb1}. The states in the $\rho$-mode have higher energies and larger rms radii than those in the $\lambda$-mode, which stems from the HQD mechanism~\cite{Li241}.
The radial and orbital excitations compete with each other, which preliminarily determines the overall structure of the excitation spectrum as shown in Fig.~\ref{fig2}.

For the low-lying ($1S$-, $2S$-, $1P$- and $1D$-wave) states, the rms radii are generally less than 0.9 fm. They have their compact shapes and could be stable. For each of the high-lying ($3S$-, $2P$- and $1F$-wave) states in Table~\ref{tb1}, however, there is always an rms radius greater than 0.9 fm. Note that for the states with masses greater than 3200 MeV, their radii are around 1.0 fm.
A baryon with a larger radius has a lower stability. So, the very high-lying excited states such as the $1F$-wave states could be observed with a very rare opportunity.

2) Off-diagonal matrix elements

Table~\ref{tb2} shows the contribution of each spin-dependent Hamiltonian term to the off-diagonal matrix elements. One can find that the contributions to the off-diagonal matrix elements from the interactions between 1 and 2 quarks are equal to 0 apart from the three-body spin-orbit terms. In most cases, the matrix elements of $\hat{H}^{tens}_{23(31)}$ and $\hat{H}^{cont}_{23(31)}$ always partially cancel each other out, and the same is true for those of $\hat{H}^{so(v)2B}_{23(31)}$ and $\hat{H}^{so(s)2B}_{23(31)}$, as well as for those of $\hat{H}^{so(v)3B}_{23(31)}$ and $\hat{H}^{so(s)3B}_{23(31)}$. Based on the previous studies~\cite{F402,Li262} and this work, it can be seen that the partial cancellation among the matrix elements of the spin-dependent terms is a common property in the configuration mixing of the baryon system.

There are several off-diagonal matrix elements having values over 20 MeV as shown in Table~\ref{tb2}, where the contribution of the three-body spin-orbit terms $\hat{H}^{so3B}$ to these off-diagonal matrix elements is dominant. This demonstrates their significance with respect to the mixing effect. Moreover, the tensor interaction $\hat{H}^{tens}$ is the only term that its matrix element is not vanishing in all the off-diagonal matrix elements. In most cases, this matrix element has the opposite effect on the total off-diagonal matrix element.

So, all of the spin-dependent interactions in Eq. (\ref{eq3}) make contributions to the off-diagonal matrix elements. The loss of any one will lead to significantly different calculation results.

3) Mixing effect and decoupling phenomenon

The mixing effect and its influence on the energy level shifts are presented in Table~\ref{tb3}. In the $\frac{1}{2}^{-}$ and $\frac{3}{2}^{-}$ subspaces, most of the energy level shifts caused by the mixing effect are within the range of 7 to 19 MeV. This has a significant impact on the fine structure, because these shifts and the energy level gaps within the fine structure share the same order of magnitude.

By comparing the matrices in the $\frac{3}{2}^{-}$ and $\frac{5}{2}^{-}$ subspaces in Tables~\ref{tb3} and~\ref{tb4} respectively, it is found that the influence of the $1F_{\lambda}$-wave states on the energy level shifts of the $P$-wave states is less than 0.1 MeV and can be disregarded. The same is true for the $1D_{\lambda}$-wave states as discussed in Ref.~\cite{Li262}. Therefore, it can be concluded that the influence of the higher $nL$-wave groups on the concerned configuration mixing is insignificant and negligible.

In practice, the mixing effect between the $1P$- and $2P$-wave groups is also very small. The evidence is as follows. In Fig.~\ref{fig2}, after taking into account the mixing between the $1P$- and $2P$-wave groups, the energy levels of the baryon states in the $1P$-wave group are 2759, 2782, 2798, 2818, 2835, 2863 and 2886 MeV. While, they are 2759, 2783, 2799, 2818, 2837, 2863 and 2887 MeV in Ref.~\cite{Li262} where the mixing effect was taken into account just within the $1P$-wave group. The difference between the two sets of data is no more than 2 MeV. So, under a lower precision requirement, the mixing effect only needs to be considered within each $nL$-wave group. In other words, the approximate decoupling phenomenon emerges between different $nL$-wave groups. In this case, the radial quantum number $n$ and orbital angular momentum $L$ can be approximately regarded as good quantum numbers. In fact, $n$ and $L$ determine the energy level structure (shell structure) as demonstrated in Fig.~\ref{fig2}.

4) Fine structures

The predicted $\Sigma_{c}$ baryon states labeled with $(mass)J^{P}$ are displayed in Fig.~\ref{fig2}. Due to the mixing effect, the orbital angular momentum quantum number $L$ is no longer a good quantum number strictly. In Fig.~\ref{fig2}, however, the baryon states are still plotted in groups with different $nL$ values according to the above discussion. This helps clearly identify which excited state is the main contributor to a baryon state. At the same time, the fine structures are clearly displayed in the four $nL$-wave groups, i.e., the $1P$-, $2P$-, $1D$- and $1F$-wave fine structures. Here, for instance, the $1P$-wave fine structure means the fine structure in the energy level region of the $1P$-wave group.

The fine structure results from the energy level splitting, shift and configuration mixing caused by the spin-dependent interactions~\cite{Li26,Li262}. As is shown in Fig.~\ref{fig2}, the energy level gaps within each fine structure group evolve along with $L$. As $L$ increases, the gaps gradually become smaller. The level gaps in the $1D$-wave fine structure are smaller than those in the $1P$-wave fine structure. In the $1F$-wave fine structure, the gaps become very small and the energy level quasi-degeneracy occurs. In the $2P$-wave fine structure, the $(3176)\frac{1}{2}^{-}$ and $(3176)\frac{3}{2}^{-}$ states also exhibit the energy level degeneracy. This poses a big difficulty for the identification of the $\Sigma_{c}(3200)^{0}$.

5) Assignment of the new $\Sigma_{c}$ baryons

Very recently, the LHCb collaboration reported three $\Sigma_{c}$ baryons with two groups of solutions~\cite{LHCb262}. In Group A, their masses are $m_{\Sigma_{c}(2800)^{0}}$ = 2819 $\pm$ 8 MeV, $m_{\Sigma_{c}(2900)^{0}}$ = 2908 $\pm$ 13 MeV and $m_{\Sigma_{c}(3200)^{0}}$ = 3186 $\pm$ 20 MeV. In Group B, their masses are $m_{\Sigma_{c}(2800)^{0}}$ = 2848 $\pm$ 10 MeV, $m_{\Sigma_{c}(2900)^{0}}$ = 2914 $\pm$ 11 MeV and $m_{\Sigma_{c}(3200)^{0}}$ = 3190 $\pm$ 16 MeV. We show these baryons from Group A in Fig.~\ref{fig2}. A nice match between the data and the predicted ones is clearly demonstrated. The mass value of the $\Sigma_{c}(2800)^{0}$ measured by the LHCb collaboration is different from the one collected in the PDG~\cite{F201}. This suggests they may not be the same baryon state. So, we temporarily mark the $\Sigma_{c}(2800)^{0}$ reported in~\cite{LHCb262} as $\Sigma_{c}(2819)^{0}$ in this work and assign it to the $(2818)\frac{3}{2}^{-}$ baryon state as shown in Fig.~\ref{fig2}. In addition, the central mass values of the $\Sigma_{c}(2800)^{0}$ given by Group A and B differ significantly. It implies that the experiment cannot specify this baryon state. If the data from Group B are considered, the mass of the $\Sigma_{c}(2800)^{0}$ in~\cite{LHCb262} is very close to that of the $\Sigma_{c}(2846)^{0}$ state earlier observed by the BABAR collaboration~\cite{F209}. Its $J^{P}$ values might be $\frac{1}{2}^{-}$ or $\frac{5}{2}^{-}$.

Considering its mass value, the $\Sigma_{c}(2900)^{0}$ is the best candidate of the predicted $(2913)\frac{1}{2}^{+}$ baryon state in the $2S$-wave state energy region as shown in Fig.~\ref{fig2}, which is consistent with the assignment of~\cite{Hiyama26} based on the data from Group B. In Refs.~\cite{Luo26,Zhang26}, however, the $\Sigma_{c}(2900)^{0}$ was interpreted as a $P$-wave state. In this work, we still assign the $\Lambda_{c}(2910)^{+}$ as a member of the $\Sigma_{c}$ family as discussed in Ref.~\cite{Li262}. If this judgment is correct, the $\Lambda_{c}(2910)^{+}$ should be named as $\Sigma_{c}(2910)^{+}$, an isospin partner of the $\Sigma_{c}(2900)^{0}$.

It is a bit difficult to make a precise assignment for the $\Sigma_{c}(3200)^{0}$ as mentioned above. Ref.~\cite{Hiyama26} considered the $\Sigma_{c}(3200)^{0}$ to be a baryon state with $J^{P}$ = $\frac{3}{2}^{-}$. In Ref.~\cite{Chen26}, however, it was interpreted as a molecular state with $J^{P}$ = $\frac{1}{2}^{-}$. In Fig.~\ref{fig2}, one can find that the measured mass error of the $\Sigma_{c}(3200)^{0}$ covers the mass range of the three predicted baryon states $(3176)\frac{1}{2}^{-}$, $(3176)\frac{3}{2}^{-}$ and $(3192)\frac{3}{2}^{-}$. So, its $J^{P}$ values might be $\frac{1}{2}^{-}$ or $\frac{3}{2}^{-}$. Alternatively, the $\Sigma_{c}(3200)^{0}$ is not a single state but likely an overlap of the predicted $(3176)\frac{1}{2}^{-}$, $(3176)\frac{3}{2}^{-}$ and $(3192)\frac{3}{2}^{-}$ states. Because the measured $\Sigma_{c}(3200)^{0}$ state is located in the predicted fine structure energy region where the energy level gaps are very small, the precise identification of the $\Sigma_{c}(3200)^{0}$ requires a higher experimental resolution.

Given the significant discrepancy between Group A and Group B of solutions from the experiment, it is necessary to examine the original data. The original distribution of the $m_{\Lambda_{c}^{+}\pi^{-}}$ invariant masses was displayed in the left panels of the figure 2 in Ref.~\cite{LHCb262}. There are two wide peaks around the energy regions of 2820 and 3200 MeV, respectively. The same structure was also observed earlier by the BABAR collaboration~\cite{F209,WKL22}. From Fig.~\ref{fig2}, it can be seen that these two wide peaks are precisely located within the energy regions of the $1P$- and $2P$-wave fine structures, respectively. Their predicted central mass values in the two energy regions are 2822 ($\frac{2759+2886}{2}$ from the $1P$-wave group) and 3210 MeV ($\frac{3153+3267}{2}$ from the $2P$-wave group), respectively. Therefore, we speculate that there might be more $\Sigma_{c}$ baryon states hidden around these two peaks. If the resolution is high enough, it should be possible to detect the two groups of fine structures in experiment.

\section{concluding remarks}\label{sec4}

The recently observed $\Sigma_{c}$ baryon states by the LHCb collaboration exceed the range of the low-lying $\Sigma_{c}$ baryon spectrum provided in our previous work.
For explaining these new baryon states, the calculated excitation spectrum of the $\Sigma_{c}$ baryons should be extended up to the energy region of high-lying excited states.
In practice, we proposed the truncation convention to cut off the excluded $nL$-wave groups and collect the included ones, i.e., the higher $nL$-wave groups should be excluded as much as possible in calculation, unless they are necessary for explaining the experimental data. Then, the $1P_{\lambda}$-, $1P_{\rho}$-, $2P_{\lambda}$-, $2P_{\rho}$- and $1F_{\lambda}$-wave states were included in the calculation according to the HQD mechanism.

The energy levels and rms radii of the included states were analyzed. It is shown that the rms radii and excited energy levels are strongly influenced by $n$ and $L$ which determine the energy level structure and less affected by $J$ and $j$. The HQD mechanism is responsible for the fact that the states in the $\rho$-mode have higher energies and larger radii than those in the $\lambda$-mode. The low-lying states have their rms radii less than 0.9 fm. They have compact shapes and could be stable. However, for the states with masses greater than 3200 MeV, their radii are around 1.0 fm, suggesting they are not so compact and might be less stable. Therefore, the observed $\Sigma_{c}(2800)^{0}$ and $\Sigma_{c}(2900)^{0}$ have compact shapes and are stable. Nevertheless, the observed $\Sigma_{c}(3200)^{0}$ would not have a compact shape and might be loosely bound state, therefore less stable.

By taking into account the mixing effect between these states, the negative-parity $\Sigma_{c}$ baryon states were obtained. Together with our previously predicted positive-parity $\Sigma_{c}$ baryon states, the updated $\Sigma_{c}$ baryon spectrum has been established in this work.
The calculation shows that the spin-dependent interactions dominate the energy level splitting and shift, the configuration mixing and the formation of the fine structures. Any type of spin-dependent interaction can not be ignored in calculations. The tensor terms and three-body spin-orbit terms are discussed especially due to their important roles in the mixing effect.

This work has extended the $\Sigma_{c}$ spectrum to the high-lying excited states by including the $2P$- and $1F$-wave states, which provides an opportunity to discover some new properties of the spectral structure. By analyzing the mixing effect, it is found that the approximate decoupling phenomenon occurs in different $nL$-wave groups. This means that the high-lying $nL$-wave groups can be ignored safely in calculations with a relatively lower precision requirement if one focuses on the low-lying excited states. This also implies that the truncation convention discussed in the calculation strategy is reasonable, which is of great significance for further improving the calculation scheme for all baryon spectroscopy. In this case, the radial quantum number $n$ and orbital angular momentums $L$ can be approximately regarded as good quantum numbers. Based on this consideration, the baryon states are plotted in groups with quantum numbers $n$ and $L$ for clearly displaying the main contributor of the $nL$ excited states to a specific baryon state. $n$ and $L$ determine the shell structure of the updated $\Sigma_{c}$ baryon spectrum as shown in Fig.2 where the four wave groups and the fine structure in each group are displayed. In the very high-lying wave groups such as the $1F$-wave group, the energy level gaps become small and the energy level quasi-degeneracy occurs. This poses a big challenge for identifying the associated baryon states.

The assignments of the newly observed $\Sigma_{c}$ baryons were proposed in this work. The measured $\Sigma_{c}(2800)^{0}$ by the LHCb collaboration is more likely a predicted $\Sigma_{c}(2819)^{0}$ baryon state with $\frac{3}{2}^{-}$ spin-parity values based on Group A data. If the data from Group B are used, however, the measured $\Sigma_{c}(2800)^{0}$ is very close to the early observed $\Sigma_{c}(2846)^{0}$ state. Its $J^{P}$ values might be $\frac{1}{2}^{-}$ or $\frac{5}{2}^{-}$. The $\Sigma_{c}(2900)^{0}$ might be assigned as a $\frac{1}{2}^{+}$ baryon state in the energy region of the $2S$-wave states. While, the measured mass error of the $\Sigma_{c}(3200)^{0}$ covers the mass range of the three predicted baryon states in the energy region of the $2P$-wave states. So, its $J^{P}$ values might be $\frac{1}{2}^{-}$ or $\frac{3}{2}^{-}$.

Given the significant discrepancy between Group A and Group B of solutions from the experiment, we investigated the original data. By analyzing the distribution of the $m_{\Lambda_{c}^{+}\pi^{-}}$ invariant masses in the experiments and the calculated results in this work, it can be seen that these two observed wide peaks are precisely located within the predicted energy regions of the $1P$- and $2P$-wave fine structures, respectively. So, we suggest more precise experiments to explore the two wide peaks in the energy regions around 2822 and 3210 MeV where one might find more $\Sigma_{c}$ baryon states.

\begin{large}
\section*{Acknowledgements}
\end{large}

 This research was supported by the Natural Science Foundation of Guizhou Province-ZK[2024](General Project)650, the National Natural Science Foundation of China (Grant Nos. 11675265, 12575083), the Continuous Basic Scientific Research Project (Grant No. WDJC-2019-13) and the Leading Innovation Project (Grant No. LC 192209000701).

\begin{large}
\section*{Appendix}
\end{large}

\begin{table*}[htbp]
\begin{ruledtabular}\caption{Hamiltonian expectation values $\langle H\rangle$ (in MeV) and root-mean-square radii (in fm) of the included radially and orbitally excited states $nL(J^{P})_{j}$ in the $\Sigma_{c}$ family.}
\begin{tabular}{c c c c c| c c c c c c c c c c c c c c}
\label{tb1}
States & $\langle r^{2}_{\rho}\rangle^{1/2}$ & $\langle r^{2}_{\lambda}\rangle^{1/2}$ & $\langle H\rangle$ & & States & $\langle r^{2}_{\rho}\rangle^{1/2}$ & $\langle r^{2}_{\lambda}\rangle^{1/2}$ & $\langle H\rangle$ \\\hline
$1S(\frac{1}{2}^{+})_{1}$ &0.611 &0.450 &2456.24 && $1S(\frac{3}{2}^{+})_{1}$ &0.645 &0.493 &2533.92 \\
$2S(\frac{1}{2}^{+})_{1}$ &0.841 &0.732 &2913.44 && $2S(\frac{3}{2}^{+})_{1}$ &0.837 &0.783 &2967.28 \\
$3S(\frac{1}{2}^{+})_{1}$ &0.945 &0.718 &3109.47 && $3S(\frac{3}{2}^{+})_{1}$ &0.992 &0.696 &3126.65 \\ \hline
$1P_{\lambda}(\frac{1}{2}^{-})_{0}$ &0.658 &0.640 &2773.06 && $2P_{\lambda}(\frac{1}{2}^{-})_{0}$ &0.807 &0.938 &3169.48 \\
$1P_{\lambda}(\frac{1}{2}^{-})_{1}$ &0.662 &0.647 &2778.02 && $2P_{\lambda}(\frac{1}{2}^{-})_{1}$ &0.807 &0.947 &3167.38 \\
$1P_{\lambda}(\frac{3}{2}^{-})_{1}$ &0.670 &0.672  &2810.40 && $2P_{\lambda}(\frac{3}{2}^{-})_{1}$ &0.801 &0.980 &3191.91\\
$1P_{\lambda}(\frac{3}{2}^{-})_{2}$ &0.678 &0.688 &2816.13 && $2P_{\lambda}(\frac{3}{2}^{-})_{2}$ &0.803 &0.996 &3185.97\\
$1P_{\lambda}(\frac{5}{2}^{-})_{2}$ &0.689  &0.731  &2862.97 && $2P_{\lambda}(\frac{5}{2}^{-})_{2}$ &0.798 &1.053 &3224.00\\
$1P_{\rho}(\frac{1}{2}^{-})_{1}$ &0.857 &0.486 &2828.13 && $2P_{\rho}(\frac{1}{2}^{-})_{1}$ &1.117 &0.713 &3221.18\\
$1P_{\rho}(\frac{3}{2}^{-})_{1}$ &0.875 &0.505 &2877.37 && $2P_{\rho}(\frac{3}{2}^{-})_{1}$ &1.116 &0.736 &3255.54\\ \hline
$1D_{\lambda}(\frac{1}{2}^{+})_{1}$ &0.683 &0.817 &3048.14 && $1F_{\lambda}(\frac{3}{2}^{-})_{2}$ &0.692 &0.995 &3295.88\\
$1D_{\lambda}(\frac{3}{2}^{+})_{1}$ &0.684 &0.834 &3062.98 && $1F_{\lambda}(\frac{5}{2}^{-})_{2}$ &0.690 &1.011 &3302.21\\
$1D_{\lambda}(\frac{3}{2}^{+})_{2}$ &0.690 &0.846 &3061.57 &&  $1F_{\lambda}(\frac{5}{2}^{-})_{3}$ &0.697 &1.025 &3296.93\\
$1D_{\lambda}(\frac{5}{2}^{+})_{2}$ &0.691 &0.871 &3082.51 && $1F_{\lambda}(\frac{7}{2}^{-})_{3}$ &0.696 &1.043  &3306.43\\
$1D_{\lambda}(\frac{5}{2}^{+})_{3}$ &0.700 &0.891 &3076.68 && $1F_{\lambda}(\frac{7}{2}^{-})_{4}$ &0.705 &1.063 &3295.33\\
$1D_{\lambda}(\frac{7}{2}^{+})_{3}$ &0.702 &0.923 &3101.93 && $1F_{\lambda}(\frac{9}{2}^{-})_{4}$ &0.704 &1.086  & 3307.04\\
\end{tabular}
\end{ruledtabular}
\end{table*}

\begin{table*}[htbp]
\begin{ruledtabular}\caption{Off-diagonal matrix elements (in MeV) of each spin-dependent Hamiltonian term in the $\frac{1}{2}^{-}$, $\frac{3}{2}^{-}$, $\frac{5}{2}^{-}$ and $\frac{7}{2}^{-}$ subspaces, respectively. }
\begin{tabular}{c c c c c c c c c c c c c c c c c c c}
\label{tb2}
 Matrix elements & $\tilde{H}^{tens}_{12}$ & $\tilde{H}^{tens}_{23(31)}$ & $\tilde{H}^{cont}_{12}$ & $\tilde{H}^{cont}_{23(31)}$  & $\tilde{H}^{so(v)2B}_{12}$ & $\tilde{H}^{so(v)2B}_{23(31)}$ & $\tilde{H}^{so(s)2B}_{12}$ & $\tilde{H}^{so(s)2B}_{23(31)}$  & $\tilde{H}^{so(s)3B}_{12}$ & $\tilde{H}^{so(s)3B}_{23(31)}$  & $\tilde{H}^{so(v)3B}_{12}$ & $\tilde{H}^{so(v)3B}_{23(31)}$& Total  \\\hline
$\langle1P_{\lambda}(\frac{1}{2}^{-})_{0}|\hat{O}|1P_{\lambda}(\frac{1}{2}^{-})_{1}\rangle$
               & 0 & 6.15    & 0 & -6.80  & 0 & -11.33   &  0 & 2.73   &0 &-0.28 &0 & 3.86& -11.33  \\
$\langle1P_{\lambda}(\frac{1}{2}^{-})_{0}|\hat{O}|2P_{\lambda}(\frac{1}{2}^{-})_{1}\rangle$
               & 0 & -1.35   &  0 & 4.49   &  0 & 5.25    &   0 & -0.58     & 0 &0.05  & 0 & -1.39 & 12.93  \\
$\langle1P_{\lambda}(\frac{1}{2}^{-})_{0}|\hat{O}|1P_{\rho}(\frac{1}{2}^{-})_{1}\rangle$
               &  0 & -5.62     &  0 & 3.96   &  0 & 0     &   0 & 0     & 0 &0  & 0 & 0 & -3.31  \\
$\langle1P_{\lambda}(\frac{1}{2}^{-})_{0}|\hat{O}|2P_{\rho}(\frac{1}{2}^{-})_{1}\rangle$
               &  0 & 0.31     &  0 & -1.77   &  0 & 0     &   0 & 0     & 0 &0  & 0 & 0 & -2.91  \\
$\langle2P_{\lambda}(\frac{1}{2}^{-})_{0}|\hat{O}|1P_{\lambda}(\frac{1}{2}^{-})_{1}\rangle$
               &  0 & -1.32     &  0 & 4.34   &  0 & 5.00     &   0 & -0.51    & 0 &0.05  & 0 & -1.29 & 12.53  \\
$\langle2P_{\lambda}(\frac{1}{2}^{-})_{0}|\hat{O}|2P_{\lambda}(\frac{1}{2}^{-})_{1}\rangle$
               &  0 & 4.04     &  0 & -5.54   &  0 & -8.11     &   0 & 2.20     & 0 &-0.20  & 0 & 2.43 & -10.35  \\
$\langle2P_{\lambda}(\frac{1}{2}^{-})_{0}|\hat{O}|1P_{\rho}(\frac{1}{2}^{-})_{1}\rangle$
               &  0 & 1.06     &  0 & -2.13   &  0 & 0     &   0 & 0     & 0 &0   & 0 & 0  & -2.14  \\
$\langle2P_{\lambda}(\frac{1}{2}^{-})_{0}|\hat{O}|2P_{\rho}(\frac{1}{2}^{-})_{1}\rangle$
               &  0 & -3.98     &  0 & 2.84   &  0 & 0     &   0 & 0     & 0 &0   & 0 & 0  & -2.28  \\
$\langle1P_{\lambda}(\frac{1}{2}^{-})_{1}|\hat{O}|1P_{\rho}(\frac{1}{2}^{-})_{1}\rangle$
               &  0 & -3.61     &  0 & 5.55   &  0 & -7.06    &   0 & 1.82    & -19.19 &-12.91  & -11.33 & 20.34 & -22.26  \\
$\langle1P_{\lambda}(\frac{1}{2}^{-})_{1}|\hat{O}|2P_{\rho}(\frac{1}{2}^{-})_{1}\rangle$
               &  0 & 0.98     &  0 & -2.45   &  0 & 0.19    &   0 & 0.72    & 4.31 &-5.14  & 8.14 & -5.13 & -9.19  \\
$\langle2P_{\lambda}(\frac{1}{2}^{-})_{1}|\hat{O}|1P_{\rho}(\frac{1}{2}^{-})_{1}\rangle$
               &  0 & 1.46     &  0 & -3.09   &  0 & 1.14    &   0 & 0.39    & -6.17 &-1.29  & 0.84 & -5.96 & -20.04  \\
$\langle2P_{\lambda}(\frac{1}{2}^{-})_{1}|\hat{O}|2P_{\rho}(\frac{1}{2}^{-})_{1}\rangle$
               &  0 & -1.83     &  0 & 4.01   &  0 & -4.12    &   0 & 1.52    & -16.87 &-11.01  & -9.39 & 12.80 & -23.52  \\ \hline
$\langle1P_{\lambda}(\frac{3}{2}^{-})_{1}|\hat{O}|1P_{\lambda}(\frac{3}{2}^{-})_{2}\rangle$
               &  0 & -4.63     &  0 & 4.65   &  0 & 8.15    &   0 & -2.08    & 0 &0.20  & 0 & -2.76 & 7.09  \\
$\langle1P_{\lambda}(\frac{3}{2}^{-})_{1}|\hat{O}|2P_{\lambda}(\frac{3}{2}^{-})_{2}\rangle$
               &  0 & 0.71     &  0 & -3.31   &  0 & -4.10    &   0 & 0.50    & 0 &-0.05  & 0 & 1.16 & -10.21  \\
$\langle1P_{\lambda}(\frac{3}{2}^{-})_{1}|\hat{O}|1P_{\rho}(\frac{3}{2}^{-})_{1}\rangle$
               &  0 & 2.76     &  0 & -2.57   &  0 & -6.87    &   0 & 1.84    & -19.00 &-13.15  & -10.82 & 19.48 & -26.82  \\
$\langle1P_{\lambda}(\frac{3}{2}^{-})_{1}|\hat{O}|2P_{\rho}(\frac{3}{2}^{-})_{1}\rangle$
               &  0 & -0.31     &  0 & 1.16   &  0 & 0.28    &   0 & 0.70    & 4.74 &-5.13  & 7.96 & -5.08 & -4.05  \\
$\langle1P_{\lambda}(\frac{3}{2}^{-})_{1}|\hat{O}|1F_{\lambda}(\frac{3}{2}^{-})_{2}\rangle$
               &  0 & 3.26     &  0 & 0   &  0 & 0    &   0 & 0    & 0 &0  & 0 & 0 & 6.52  \\
$\langle2P_{\lambda}(\frac{3}{2}^{-})_{1}|\hat{O}|1P_{\lambda}(\frac{3}{2}^{-})_{2}\rangle$
               &  0 & -0.26     &  0 & 3.10   &  0 & 3.72    &   0 & -0.38    & 0 &0.04  & 0 & -1.01 & 10.41  \\
$\langle2P_{\lambda}(\frac{3}{2}^{-})_{1}|\hat{O}|2P_{\lambda}(\frac{3}{2}^{-})_{2}\rangle$
               &  0 & 2.86     &  0 & -4.06   &  0 & -6.09    &   0 & 1.70    & 0 &-0.15  & 0 & 1.81 & -7.86  \\
$\langle2P_{\lambda}(\frac{3}{2}^{-})_{1}|\hat{O}|1P_{\rho}(\frac{3}{2}^{-})_{1}\rangle$
               &  0 & 0.74     &  0 & -1.50   &  0 & -1.31    &   0 & -0.33    & 5.76 &1.09  & -0.78 & 6.07 & 14.49  \\
$\langle2P_{\lambda}(\frac{3}{2}^{-})_{1}|\hat{O}|2P_{\rho}(\frac{3}{2}^{-})_{1}\rangle$
               &  0 & -1.60     &  0 & 1.91   &  0 & 3.97    &   0 & -1.48    & 16.82 &11.01  & 9.35 & -12.40 & 28.97  \\
$\langle2P_{\lambda}(\frac{3}{2}^{-})_{1}|\hat{O}|1F_{\lambda}(\frac{3}{2}^{-})_{2}\rangle$
               &  0 & -1.17     &  0 & 0   &  0 & 0    &   0 & 0    & 0 &0  & 0 & 0 & -2.33  \\
$\langle1P_{\lambda}(\frac{3}{2}^{-})_{2}|\hat{O}|1P_{\rho}(\frac{3}{2}^{-})_{1}\rangle$
               &  0 & -5.93  &  0 & 5.61  &  0 & 0     &   0 & 0   & 0 &0  & 0 & 0 & -0.64  \\
$\langle1P_{\lambda}(\frac{3}{2}^{-})_{2}|\hat{O}|2P_{\rho}(\frac{3}{2}^{-})_{1}\rangle$
               &  0 & -0.97  &  0 & -2.46  &  0 & 0  &   0 & 0   & 0 &0  & 0 & 0 & -6.87  \\
$\langle1P_{\lambda}(\frac{3}{2}^{-})_{2}|\hat{O}|1F_{\lambda}(\frac{3}{2}^{-})_{2}\rangle$
               &  0 & 1.47  &  0 & 0  &  0 & 0 &   0 & 0  & 0 &0  & 0 & 0 & 2.93  \\
$\langle2P_{\lambda}(\frac{3}{2}^{-})_{2}|\hat{O}|1P_{\rho}(\frac{3}{2}^{-})_{1}\rangle$
               &  0 & 0.51  &  0 & -3.50   &  0 & 0  &   0 & 0   & 0 &0  & 0 & 0 & -5.98  \\
$\langle2P_{\lambda}(\frac{3}{2}^{-})_{2}|\hat{O}|2P_{\rho}(\frac{3}{2}^{-})_{1}\rangle$
               &  0 & -4.26  &  0 & 4.24   &  0 & 0  &   0 & 0     & 0 &0  & 0 & 0 & -0.04  \\
$\langle2P_{\lambda}(\frac{3}{2}^{-})_{2}|\hat{O}|1F_{\lambda}(\frac{3}{2}^{-})_{2}\rangle$
               &  0 & 0.46  &  0 & 0  &  0 & 0   &   0 & 0  & 0 &0  & 0 & 0 & 0.93  \\
$\langle1P_{\rho}(\frac{3}{2}^{-})_{1}|\hat{O}|1F_{\lambda}(\frac{3}{2}^{-})_{2}\rangle$
               &  0 & 1.05  &  0 & 0   &  0 & 0  &   0 & 0   & -2.05 &0  & -1.17 & 0 & -1.12  \\
$\langle2P_{\rho}(\frac{3}{2}^{-})_{1}|\hat{O}|1F_{\lambda}(\frac{3}{2}^{-})_{2}\rangle$
               &  0 & 0.38  &  0 & 0   &  0 & 0    &   0 & 0   & 1.18 &0  & 1.17 & 0 & 3.11  \\ \hline
$\langle1P_{\lambda}(\frac{5}{2}^{-})_{2}|\hat{O}|1F_{\lambda}(\frac{5}{2}^{-})_{2}\rangle$
               &  0 & -0.94  &  0 & 0   &  0 & 0  &   0 & 0   & 0 &0  & 0 & 0 & -1.88 \\
$\langle1P_{\lambda}(\frac{5}{2}^{-})_{2}|\hat{O}|1F_{\lambda}(\frac{5}{2}^{-})_{3}\rangle$
               &  0 & 4.13    &  0 & 0   &  0 & 0  &   0 & 0   & 0 &0 & 0 & 0 & 8.26  \\
$\langle2P_{\lambda}(\frac{5}{2}^{-})_{2}|\hat{O}|1F_{\lambda}(\frac{5}{2}^{-})_{2}\rangle$
               &  0 & -0.21  &  0 & 0   &  0 & 0  &   0 & 0  & 0 &0  & 0 & 0 & -0.42  \\
$\langle2P_{\lambda}(\frac{5}{2}^{-})_{2}|\hat{O}|1F_{\lambda}(\frac{5}{2}^{-})_{3}\rangle$
               &  0 & 0.97  &  0 & 0   &  0 & 0  &   0 & 0  & 0 &0  & 0 & 0 & 1.95 \\
$\langle1F_{\lambda}(\frac{5}{2}^{-})_{2}|\hat{O}|1F_{\lambda}(\frac{5}{2}^{-})_{3}\rangle$
               &  0 & -2.75  &  0 & 0.76  &  0 & 2.84 &   0 & -1.80  & 0 &0.10  & 0 & -0.61 & -2.93  \\  \hline
$\langle1F_{\lambda}(\frac{7}{2}^{-})_{3}|\hat{O}|1F_{\lambda}(\frac{7}{2}^{-})_{4}\rangle$
               &  0 & -2.05  &  0 & 0.56  &  0 & 2.26     &   0 & -1.52    & 0 &0.08  & 0 & -0.47 & -2.30  \\
\end{tabular}
\end{ruledtabular}
\end{table*}

\begin{table*}[htbp]
\begin{ruledtabular}\caption{Hamiltonian matrix $\tilde{H}$ (in MeV), eigenvalues (in MeV) and eigenvectors in the $\frac{1}{2}^{-}$, $\frac{3}{2}^{-}$, $\frac{5}{2}^{-}$, $\frac{7}{2}^{-}$ and $\frac{9}{2}^{-}$ subspaces of the $\Sigma_{c}$ baryon states including the $1F$-wave states.}
\begin{tabular}{c c c c c c c c c c c c c c c c c c c}
\label{tb3}
$J^{P}$ & States & $\tilde{H}$ & Eigenvalue & Eigenvector   \\\hline
\multirow{6}{*}{$\frac{1}{2}^{-}$} &$1P_{\lambda}(\frac{1}{2}^{-})_{0}$ & \multirow{6}{*}{$ \begin{pmatrix} 2773.06 & 0 & -11.33 & 12.93 & -3.31 & -2.91\\ 0 & 3169.48 & 12.53 & -10.35 & -2.14 & -2.28\\ -11.33 & 12.53 & 2778.02 & 0 & -22.26 & -9.19\\ 12.93&-10.35 & 0 & 3167.38 & -20.04 & -23.52 \\ -3.31&-2.14 & -22.26& -20.04 & 2828.13 & 0 \\ -2.91&-2.28 & -9.19 & -23.52 & 0 & 3221.18\end{pmatrix} $}  &3230.46 & (-0.02, 0.02, -0.02, -0.36, 0.02, $\mathbf{0.93}$)\\
     & $2P_{\lambda}(\frac{1}{2}^{-})_{0}$ & &3176.45 &(-0.02, $\mathbf{0.85}$, 0.03, -0.49, 0.021, -0.21)\\
     & $1P_{\lambda}(\frac{1}{2}^{-})_{1}$ & &3153.38 &(0.03, 0.53, 0.01, $\mathbf{0.79}$, -0.05, 0.30)\\
      & $2P_{\lambda}(\frac{1}{2}^{-})_{1}$ & &2835.42 &(-0.03, -0.02, 0.36, -0.06, -$\mathbf{0.93}$, 0.00)\\
       & $1P_{\rho}(\frac{1}{2}^{-})_{1}$ & &2782.48 &($\mathbf{0.76}$, 0.02, -0.60, -0.04, -0.25, -0.01)\\
    & $2P_{\rho}(\frac{1}{2}^{-})_{1}$ & &2759.06 &(0.65, -0.02, $\mathbf{0.71}$, -0.01, 0.26, 0.02)\\ \hline
\multirow{7}{*}{$\frac{3}{2}^{-}$} &$1P_{\lambda}(\frac{3}{2}^{-})_{1}$ & \multirow{7}{*}{$ \begin{pmatrix} 2810.40 & 0&7.09&-10.21&-26.82 & -4.05& 6.52 \\ 0& 3191.91&10.41& -7.86& 14.49 &28.97&-2.33 \\ 7.09& 10.41& 2816.13& 0 &-0.64 &-6.87& 2.93 \\-10.21& -7.86&0 &3185.97&-5.98 &-0.04& 0.93 \\ -26.82&14.49&-0.64&-5.98&2877.37& 0& -1.12 \\ -4.05&28.97&-6.87&-0.04&0& 3255.54& 3.11\\ 6.52&-2.33&2.93&0.93&-1.12& 3.11& 3295.88 \end{pmatrix} $}  &3296.22 & (0.01, -0.00, 0.01, 0.01, -0.00, 0.07, $\mathbf{1.00}$)\\
     & $2P_{\lambda}(\frac{3}{2}^{-})_{1}$ & &3266.86 &(0.01, -0.37, 0.01, 0.04, -0.02, -$\mathbf{0.93}$, 0.07)\\
      & $1P_{\lambda}(\frac{3}{2}^{-})_{2}$ & &3191.80 &(-0.02, -0.54, -0.02, $\mathbf{0.80}$, -0.04, 0.24, -0.03)\\
      & $2P_{\lambda}(\frac{3}{2}^{-})_{2}$ & &3176.03 &(-0.01, $\mathbf{0.75}$, 0.03, 0.59, 0.03, -0.27, 0.02)\\
      & $1P_{\rho}(\frac{3}{2}^{-})_{1}$ & &2886.31 &(0.34, 0.04, 0.05, -0.01, -$\mathbf{0.94}$, 0.00, -0.01)\\
       & $2P_{\rho}(\frac{3}{2}^{-})_{1}$ & &2817.77 &(0.30, -0.03, $\mathbf{0.94}$, 0.01, 0.16, 0.02, -0.01)\\
    & $1F_{\lambda}(\frac{3}{2}^{-})_{2}$ & &2798.23 &(-$\mathbf{0.89}$, 0.00, 0.34, -0.03, -0.30, -0.00, 0.01)\\ \hline
\multirow{4}{*}{$\frac{5}{2}^{-}$} &$1P_{\lambda}(\frac{5}{2}^{-})_{2}$ & \multirow{4}{*}{$ \begin{pmatrix} 2862.97 & 0 & -1.88& 8.26 \\ 0& 3224.00& -0.42& 1.95 \\ -1.88& -0.42& 3302.21& -2.93\\ 8.26& 1.95& -2.93& 3296.93\end{pmatrix} $}  &3303.59 & (-0.01, -0.02, $\mathbf{0.91}$, -0.42)\\
     & $2P_{\lambda}(\frac{5}{2}^{-})_{2}$ & &3295.77 &(0.02, 0.02, 0.42, $\mathbf{0.91}$)\\
     & $1F_{\lambda}(\frac{5}{2}^{-})_{2}$ & &3223.95 &(-0.00, $\mathbf{1.00}$, 0.00, -0.03)\\
     & $1F_{\lambda}(\frac{5}{2}^{-})_{3}$ & &2862.81 &($\mathbf{1.00}$, 0.00, 0.00, -0.02)\\ \hline
\multirow{2}{*}{$\frac{7}{2}^{-}$} &$1F_{\lambda}(\frac{7}{2}^{-})_{3}$ & \multirow{2}{*}{$ \begin{pmatrix} 3306.43 & -2.30 \\ -2.30& 3295.33\end{pmatrix} $}  &3306.89 & (-$\mathbf{0.98}$, 0.20)\\
     & $1F_{\lambda}(\frac{7}{2}^{-})_{4}$ & &3294.87 &(-0.20, -$\mathbf{0.98}$)\\ \hline
$\frac{9}{2}^{-}$ & $1F_{\lambda}(\frac{9}{2}^{-})_{4}$ & (3307.04) & 3307.04 & ($\mathbf{1}$) \\
\end{tabular}
\end{ruledtabular}
\end{table*}

\begin{table*}[htbp]
\begin{ruledtabular}\caption{Hamiltonian matrix $\tilde{H}$ (in MeV), eigenvalues (in MeV) and eigenvectors in the $\frac{3}{2}^{-}$ and $\frac{5}{2}^{-}$ subspaces of the $\Sigma_{c}$ baryon states excluding the $1F$-wave states.}
\begin{tabular}{c c c c c c c c c c c c c c c c c c c}
\label{tb4}
$J^{P}$ & States & $\tilde{H}$ & Eigenvalue & Eigenvector   \\\hline
\multirow{6}{*}{$\frac{3}{2}^{-}$} &$1P_{\lambda}(\frac{3}{2}^{-})_{1}$ & \multirow{6}{*}{$ \begin{pmatrix} 2810.40 & 0&7.09&-10.21&-26.82 & -4.05 \\ 0& 3191.91&10.41& -7.86& 14.49 &28.97 \\ 7.09& 10.41& 2816.13& 0 &-0.64 &-6.87 \\-10.21& -7.86&0 &3185.97&-5.98 &-0.04 \\ -26.82&14.49&-0.64&-5.98&2877.37& 0 \\ -4.05&28.97&-6.87&-0.04&0& 3255.54 \end{pmatrix} $}  &3266.98 & (-0.01, 0.36, -0.01, -0.04, 0.01, $\mathbf{0.93}$)\\
     & $2P_{\lambda}(\frac{3}{2}^{-})_{1}$ & &3191.86 &(0.02, 0.54, 0.02, -$\mathbf{0.80}$, 0.04, -0.24)\\
      & $1P_{\lambda}(\frac{3}{2}^{-})_{2}$ & &3176.06 &(0.06, -$\mathbf{0.75}$, -0.03, -0.60, -0.03, 0.27)\\
      & $2P_{\lambda}(\frac{3}{2}^{-})_{2}$ & &2886.33 &(0.34, 0.04, 0.05, -0.01, -$\mathbf{0.94}$, 0.00)\\
      & $1P_{\rho}(\frac{3}{2}^{-})_{1}$ & &2817.81 &(0.31, -0.03, $\mathbf{0.94}$, 0.01, 0.16, 0.02)\\
       & $2P_{\rho}(\frac{3}{2}^{-})_{1}$ & &2798.27 &($\mathbf{0.89}$, -0.00, -0.34, 0.03, 0.30, 0.00)\\ \hline
\multirow{2}{*}{$\frac{5}{2}^{-}$} &$1P_{\lambda}(\frac{5}{2}^{-})_{2}$ & \multirow{2}{*}{$ \begin{pmatrix} 2862.97 & 0  \\ 0& 3224.00 \end{pmatrix} $}  &3224.00 & (0, $\mathbf{1}$)\\
     & $2P_{\lambda}(\frac{5}{2}^{-})_{2}$ & &2862.97 &($\mathbf{1}$, 0)\\
\end{tabular}
\end{ruledtabular}
\end{table*}

\end{document}